# Rapid Computational Optimization of Molecular Properties using Genetic Algorithms: Searching Across Millions of Compounds for Organic Photovoltaic Materials

*Ilana Y. Kanal,[1] Geoffrey R. Hutchison[1,2]\**

[1]Department of Chemistry, University of Pittsburgh, 219 Parkman Avenue,

Pittsburgh, Pennsylvania 15206

[2]Department of Chemical Engineering, University of Pittsburgh, Pittsburgh, PA 15261

ABSTRACT. Conjugated organic molecules represent an important area of materials chemistry for both fundamental scientific exploration and technological applications. Using a genetic algorithm to computationally screen up to ~25-50 million molecules for organic photovoltaic properties, we find that our methods find top monomers 6,000-8,000 times faster than brute force search. By testing multiple runs and establishing convergence criteria, we show the computational scaling with search space size, common molecular motifs, and discuss the reliability to choose the same molecules independent of initial data set size. We outline remaining areas of difficulty in growing to larger search spaces, potential solutions, and filtering criteria for potential organic photovoltaic materials. Efficient genetic algorithm searches promise to address a wide range of property-driven inverse design problems in chemistry.

INTRODUCTION.

Computational molecular screening is crucial to many areas of research ranging from drug discovery to discovery of new materials for many industries.[1-3] Many methods have so far been used for property-based screening, each of which has virtues and faults. A typical method attractive to many researchers is the development of large libraries of known structures, which can then be searched for molecules of interest for specific applications.[4,5] Although these libraries take significant computational time to generate, once constructed, they can be repeatedly and easily searched for molecules with particular features. For example, one such study successfully developed a library with small organic molecules for thermally activated delayed fluorescence (TADF).[5] A similar effort generated a map to search the uncharted areas of the small molecule universe, and targeting molecules that do not yet exist.[6] Similar libraries can be generated for diverse molecular species with favorable physical property values.[4]

While library generation helps with repeated screening, more typically, researchers are interested in rapidly finding new compounds with particular "best" possible properties – an inverse design problem where the figure of merit is known, but the molecular design motifs are not. New machine learning methods can provide an alternative strategy to generating of large databases. Aspuru-Guzik *et. al.* reported a generative model for efficient searching and optimization through open-ended spaces of chemical compounds. This generative model uses deep neural networks trained on hundreds of thousands of existing chemical structures to construct two coupled functions: an encoder and a decoder with their method being demonstrated for design of drug-like molecules and organic light-emitting diodes.[7] While neural network methods allows continuous optimization for molecules, inherently discrete species, so to ensure accurate results, the encoder and decoder functions require a lot of training. Another approach is the use of interpolation of property values, requiring less than 0.01% of the search space to find optimal targets.

While there are many areas of chemistry requiring rapid inverse design of molecules with improved properties, we have focused on π-conjugated organic electronic materials. Using the power of synthetic chemical tailorability, the topic has attracted fundamental scientific understanding on optoelectronic, charge transport, and other properties, as well as a huge range of potential applications.[8,9] For example, organic photovoltaics often employ the donor-acceptor approach in which electron-poor acceptor and electron-rich donor monomers are mixed to create copolymers with the desired optoelectronic properties.[10-13] Much effort has been made on finding novel monomers or side-chains to tailor the properties of the resulting oligomers and copolymers.[14]

In this work, the key question is whether genetic algorithm (GA) screening methods can be made reliable and efficient for performing discrete property-driven optimization of molecules. We have previously focused on GA methods to rapidly find optimal molecular targets for particular properties without pre-generating a library or training set. We seek to grow the search space from ~500,000 compounds,[15] ultimately to ~50 million molecules, by enlarging the pool of potential monomer "genes" from 129 to 1759, and sampling all possible sequences.[16-18] While still much smaller than all molecular space (~$10^{60}$),[19,20] by testing multiple runs and establishing a convergence criteria, we find our methods to be 6,000-8000 times faster than brute force search. We outline remaining areas of difficulty in growing to larger search spaces, potential solutions, and filtering criteria for potential organic photovoltaic materials. The promise of efficient genetic algorithm sampling of molecular properties can be applied to many other molecular search areas in the future.

**COMPUTATIONAL METHODS.**

**Table 1**. Computed ranges of ZINDO HOMO and LUMO eigenvalues for the monomers and homotetramers in each data set indicating the increasing diversity in the search pools with increased number of monomers.

| Number of Monomers | Monomer HOMO (eV) | | Monomer LUMO (eV) | | Tetramer HOMO (eV) | | Tetramer LUMO (eV) | |
|---|---|---|---|---|---|---|---|---|
| | Min | Max | Min | Max | Min | Max | Min | Max |
| 129 | -8.59 | -4.17 | -3.93 | -1.39 | -10.33 | -5.71 | -3.47 | +0.58 |
| 442 | -8.13 | -2.70 | -5.56 | +1.90 | -10.39 | -4.19 | -4.73 | +2.86 |
| 611 | -8.59 | -1.29 | -4.41 | +2.40 | -11.32 | -3.31 | -5.56 | +2.92 |
| 908 | -8.59 | -1.22 | -4.17 | +2.29 | -10.66 | -3.92 | -4.74 | +2.45 |
| 1235 | -8.59 | -1.50 | -4.41 | +2.02 | -11.00 | -3.25 | -4.51 | +2.92 |
| 1759 | -8.59 | -2.74 | -4.45 | +2.40 | -11.32 | -3.25 | -4.73 | +2.86 |

**Monomer Data Sets.** Five data sets, comprising of 129, 442, 611, 908, 1235 and 1759 monomers,(Tables S1-S41 and Figures S1-S41) were prepared for this study by selecting small monomers that are likely to be used in organic photovoltaics. Monomers were selected from literature reports or obvious synthetic modifications of conjugated monomers, to span a wide range of aromatic and conjugated species. For this study, most of the species studied contain a combination of C, H, N, O, S, F elements, and those containing Si and Se and other elements were excluded. In addition, we restricted polymerization sites to those considered most synthetically likely. A range of electron-donating and electron-with-drawing substituents were considered. The monomers span a wide range of electronic properties as shown in Table 1, with highest occupied molecular orbital (HOMO) and lowest unoccupied molecular orbital (LUMO) eigenvalues displaying multiple eV range for each (i..e, 5.8 eV range for HOMO and 6.8 eV range for LUMO eigenvaluesof monomers). For comparison, thiophene is computed to have a HOMO eigenvalue of -8.54 eV and a LUMO at -2.03 eV.

**Generation of Optimized 3D Structures.** The 3D structure of a homotetramer was generated using a multistep process starting with the SMILES string for the polymer.[21] An initial 3D structure was generated using Open Babel 2.4.0[22] (accessed through its Python interface Pybel)[23] and minimized using the MMFF94 force field (500 steps using steepest descent minimization, convergence at $1.0^{-4}$ kcal/mol). Next a weighted-rotor search (MMFF94, 100 iterations, 20 geometry optimization steps) was carried out to find a low-energy conformer. This was then further optimized using MMFF94 (500 steps). Finally, Gaussian 09[24] was used to optimize the structure using the PM6 semiempirical method.[25] The entire procedure required ~8 min per oligomer on one CPU core.

**Prediction of Electronic Structure and Optical Excitation Energies**. The energies and oscillator strengths of the 15 lowest-energy electronic transitions were calculated from the PM6-optimized geometry[25] using the ZINDO/S method[26] as implemented in Gaussian09.[24] The Python library `cclib`[27] was used to extract the molecular orbital eigenvalues, energies and oscillator strengths of the electronic transitions. The accuracy of this method was tested in our previous work[15] and produces good predictions at low computational cost.[28]

**Synthetic Accessibility.** To limit the possible search space and to concentrate on the most synthetically relevant species, we considered copolymers formed by preparing a dimer of two different monomers, followed by polymerization to make tetramers of all possible sequences (as illustrated in Scheme 1) generating 264,192 tetramers (129 monomer set), 3,118,752 tetramers (442 monomer set), 5,963,360 tetramers (611 monomer set), 13,176,869 tetramers (908 monomer set), 24,383,840 tetramers (1235 monomer set) and 49,477,152 tetramers (1759 monomer set). In addition, hexamers were tested for the data sets containing 129, 442 and 908 monomers, increasing the number of molecules screened to over 52 million compounds.

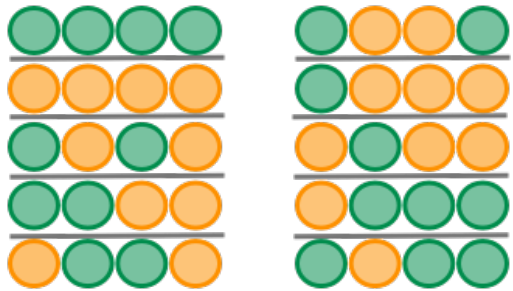

**Scheme 1**. Tetramer sequences permitted within the genetic algorithm for the combination of two monomers.

**Calculation of Energy Conversion Efficiency**. As with our previous work, the predicted power conversion efficiency was calculated as described by Scharber *et al.*[29] on the basis of the orbital energy levels and first excitation energy of the oligomer/polymer donor material relative to (PCBM). Our implementation is identical to previous work[15] and details can be found in the supporting information.

**Genetic Algorithm**. A genetic algorithm (GA) is a stochastic method for global optimization based on concepts from evolution, in which a population of "chromosomes" is optimized in successive generations by applying the evolutionary operators of crossover, mutation and selection. In our study, the chromosomes were 64 candidate co-oligomers, and successive generations minimized the deviation from the HOMO and electronic transition energy values at the point of maximum efficiency.[29] Additional weight was given for a large oscillator strength of the lowest energy excitation, since this is correlated with strong optical absorption extinction coefficients.

A key feature of our genetic algorithm implementation is the mutation operator, allowing for mutation by selecting between monomer genes with similar electronic properties. To define "similar," for each of the monomers we generated 3D structures (as described above) of the corresponding homo-oligomer of length four and carried out a ZINDO/S single-point calculation of this tetramer. Similar monomers were defined as those oligomers with similar HOMO and LUMO eigenvalues, as measured by 2D Euclidean distance. Full details of the genetic algorithm are available in the Supporting Information.

**Analysis**. Results from the genetic algorithm were analyzed using Python with `numpy`[30] and `pandas`[31] modules to generate histograms of monomers most often chosen and determine Spearman correlations and the percentage of monomers which were identical in different data sets to identify the top monomers and the number of generations needed for the convergence of the set of top monomers.

RESULTS AND DISCUSSION.

Our previous work has shown that the GA method can find candidate oligomers with high predicted efficiency by the Scharber criteria.[15] In this work, we wish to focus on understanding the speed and reliability of the stochastic GA search, particularly as it scales to larger search spaces. In particular, a key question is how long the method takes to find top candidates.

**Scaling of our GA to massive search spaces.**

Determining the number of generations required to converge the list of top performing molecules is essential to scaling a genetic algorithm to large search spaces. Ideally, data from several runs, with different starting populations of chromosomes, can provide a set of monomers in which the most frequently chosen monomers are identical in each set. Data set convergence to these sets of top monomers was determined through calculation of the Spearman rank correlation, which quantifies how well two lists are correlated in ranked order by energy. Two identically ordered lists produce a perfect Spearman correlation of +1, and an inversely ordered list provides a perfectly inverse Spearman correlation of –1. **(Scheme 2)**

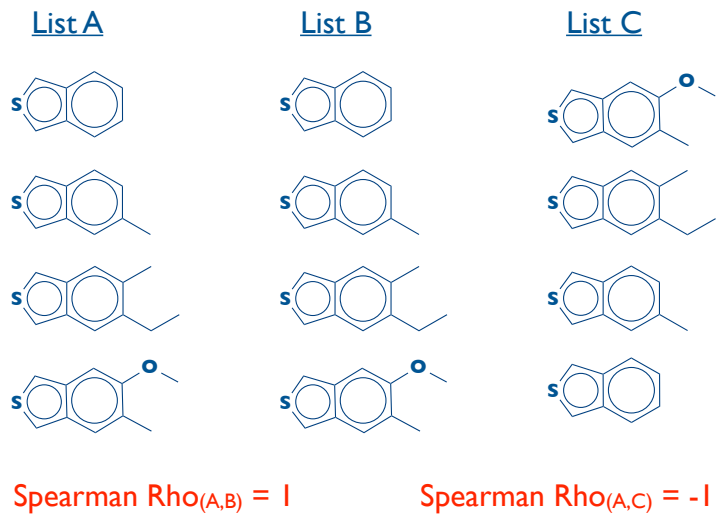

**Scheme 2**. Spearman rank correlation explained. Two identically ordered lists have a Spearman correlation of +1, while two lists which are ordered exactly opposite of each other have a Spearman correlation of -1.

For each of the five data sets described in the computational methods section, Spearman correlations (ρ) were calculated at intervals within the range of 1-100 generations. Convergence of each data set to a set of top monomers was approximated at the value where the average Spearman correlation is equal to 0.50. This cutoff was chosen due to the asymptotic scaling exhibited by the data, and therefore it would be necessary to calculate vastly more generations to achieve greater correlation (e.g., Figure S42). An analysis of this threshold value is performed below. Equations of best fit for the top 5, 10, 15, 20 and 25% of the data were determined and used to calculate the number of generations required to achieve data set convergence (Figures S42-43, Tables 42-45).

**Table 2**. The number of calculated generations to convergence of the top monomers for the top X% of the tetramer data. Convergence was calculated using the equations of fit to exceed 0.5 correlation.

| Number of Monomers | Number of Possible Compounds | 25% | 20% | 15% | 10% |
|---|---|---|---|---|---|
| 129 | 264,192 | 9.2 ± 1.1 | 8.4 ± 1.1 | 7.27 ± 0.93 | 6.76 ± 1.1 |
| 442 | 3,118,752 | 14.2 ± 2.6 | 11.7 ± 2.0 | 10.57 ± 1.94 | 10.4 ± 2.2 |
| 611 | 5,963,360 | 16.7 ± 2.0 | 20.4 ± 2.2 | 33.16 ± 2.77 | Does not converge |
| 908 | 13,176,869 | 41.2 ± 2.5 | 51.3 ± 3.6 | 59.99 ± 5.72 | 61.76 ± 6.6 |
| 1235 | 24,383,840 | 48.6 ± 4.3 | 60.7 ± 5.2 | 80.38 ± 5.26 | 97.42 ± 3.8 |
| 1759 | 49,477,152 | Does not converge | | | |

The results for the number of generations to convergence are summarized in Table 2 and show that as the data set increases in size, the number of generations needed for the convergence of the top monomer set also increases, but only modestly. For completeness, the top 25% of the hexamer data was also analyzed and results corroborate the tetramer data: the 129 monomers set converges in 5.99 ± 0.98 generations, the 442 monomer set converges in 12.86 ± 1.52 generations and the 908 monomer set converges in 57.15 ± 10.7 generations. These values are within error of the tetramer values, showing that the data set is effectively being screened, regardless of oligomer size. Using the number of generations required for each of the different sized data sets to achieve to convergence, it is possible predict the number of generations to convergence for data sets of different sizes (Figure 1).

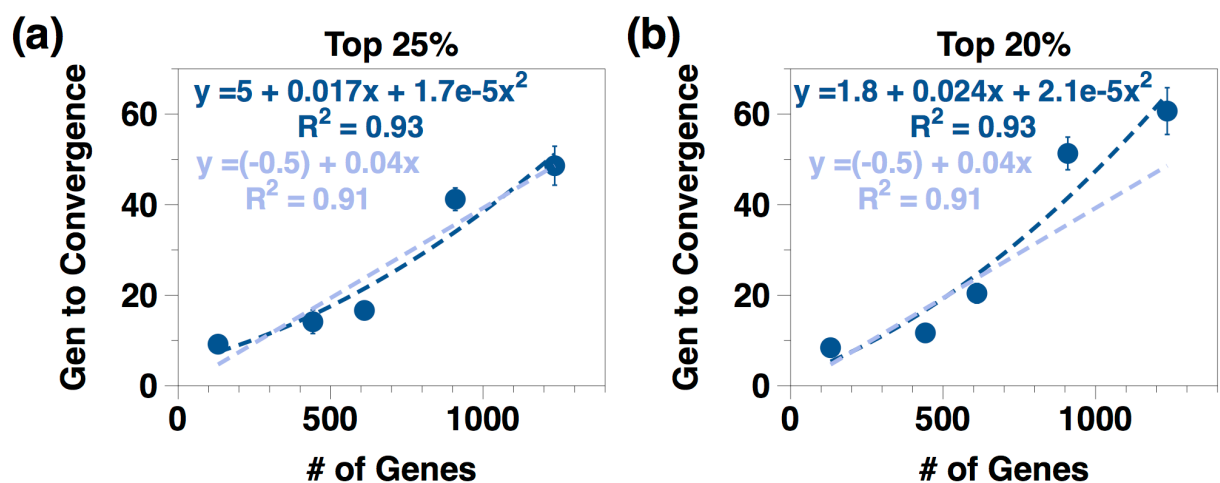

**Figure 1.** Number of generations required for top monomer convergence for different sized data sets. For both (a) the top 20% and (b) the top 25% data sets, the number of generations required for data set convergence is quadratic or linear excluding the 1759 monomer data set which did not achieve convergence.

For data sets smaller than 1235 monomers, the model is slightly quadratic or possibly linear, but at larger values (i.e., 1759) the data does not converge to a set of top monomers since the search space is evidently too large – the so-called "curse of dimensionality." Data sets examining the top 25% and 20% both show $R^2$ values for the quadratic fit of 0.93 for the prediction of the number of generations required to achieve convergence of a data set and $R^2$ values for the linear fit of 0.91, excluding the point for 1759 monomers. We hope that future work investigating larger spaces will elucidate the computational scaling of the method, but for now, it appears quasi-linear through ~25 million candidates at 1235 monomers.

The error in the number of generations to convergence, as reported in Table 2 was determined by taking the residual from the line of best fit divided by the square root of the sample size. The number of generations to convergence was then calculated for 0.5 +/- this value and the error was determined by subtracting the two new values for the generations to convergence. Unsurprisingly, as the data set increases in size, the error in the number of generations to convergence also increases, since more there is a larger space to sample, and less efficient sampling of the entire data set. Clearly this is the difficulty with 1759 monomers and

**Efficiency of our GA approach**. The discrepancy in the number of generations to convergence with 1235 and 1759 monomers demonstrates that a region exists in which the number of monomers in a data set have maximum efficiency. Each generation in our screening consists of 64 calculations (one for each "chromosome"). The fraction of the calculations which we must perform to achieve convergence for a data set of a particular size is the number of calculations to convergence (i.e., 64 calculations per generation times the number of generations to convergence) divided by the number of calculations in an exhaustive search for that data set. The speedup is therefore the reciprocal of this value or the number of calculations in an exhaustive search divided by the number of calculations to convergence (Tables S46-47).

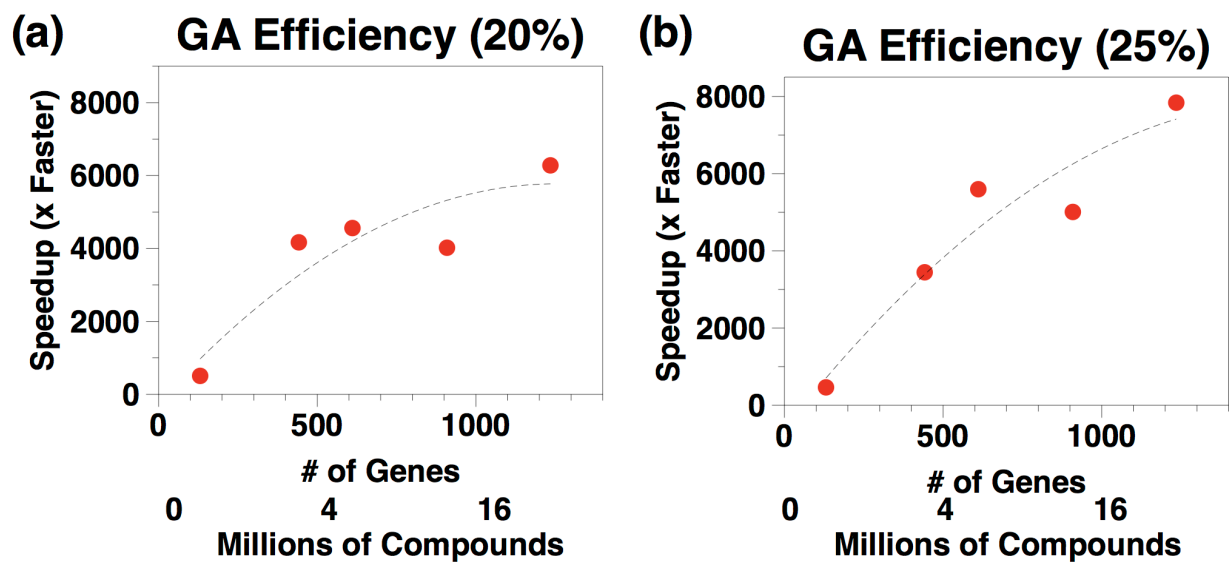

**Figure 2**. The speedup as calculated from the number of calculations performed when running the genetic algorithm compared with the number of calculations required for an exhaustive search. As shown in (a) top 20% and (b) top 25%.

While the "easiest" solution to screening millions of molecules would seem to be to screen through an exhaustive search where calculations are performed for each molecule, in reality this is a very slow and therefore costly approach. As illustrated in Figure 2, our GA converges to a set of top monomers with a speedup of ~6000x over brute force (for 1235 monomers across the top 20%), and a speedup of ~8000x (for 1235 monomers across the top 25%). Since the largest search space of ~50 million compounds and 1759 monomers did not converge, for future screening of large data sets, a tournament style approach would likely improve efficiency by dividing the search pool into several optimally sized groups (~1,000 monomers each). Each group can be screened for top monomers and then the top monomers can compete against one another to determine the overall top monomers.

**Properties of Top Candidates**

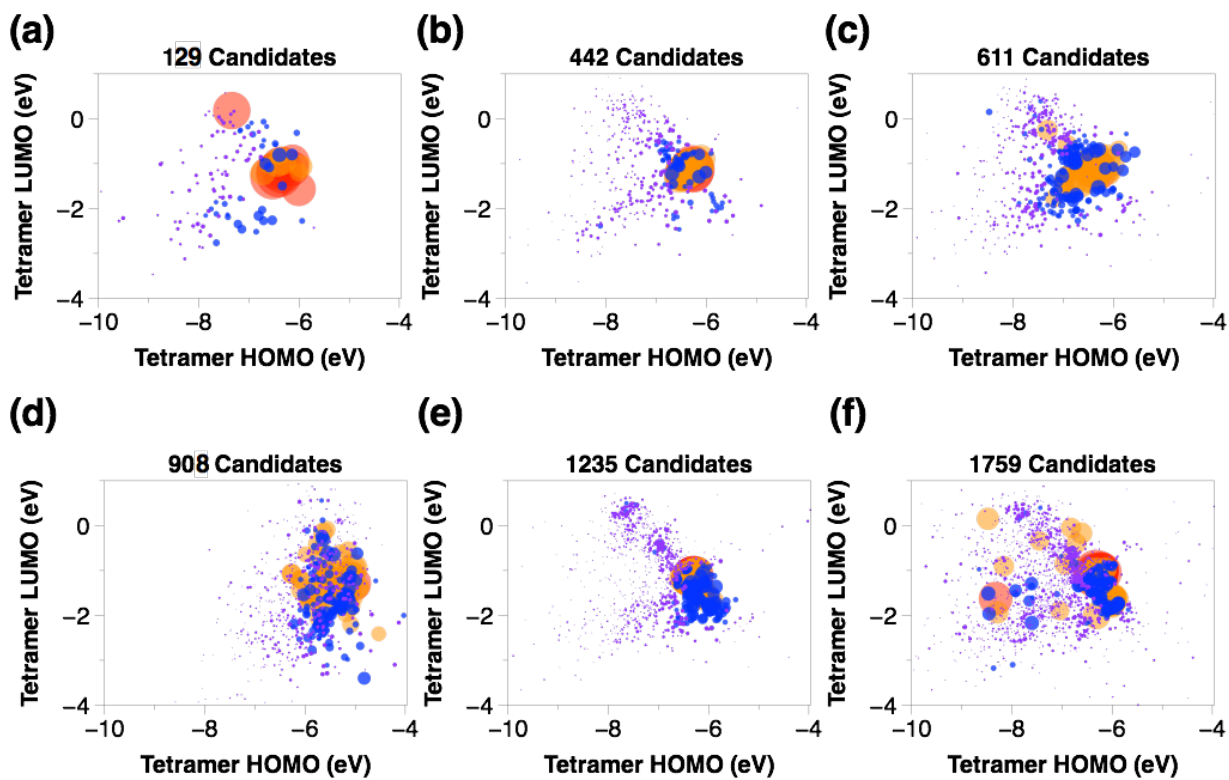

**Figure 3**. "Hotspots" of homotetramer HOMO and LUMO ZINDO eigenvalues (after 100 generations), with size and color of the spot normalized based on the frequency of occurrence of each monomer, indicates that in all data sets, tetramers in a small range of properties show top promise as candidates for solar cells. Note that the 1759 candidate set is likely not converged, but shows similar properties to other search spaces.

The initial random selection of monomers candidates yields a diverse population of HOMO and LUMO energies during the first generation in each set of data, independent of the size of the search. However, as the GA proceeds through multiple generations, HOMO and LUMO eigenvalues, which are selected in the remaining population of monomers narrows to a "hot spot" that emerges within each group of data (Figure 3). While the set of 1759 candidates is likely not converged, in most cases, with 129 monomers, 442, 611, and 1235 monomers, the shapes are comparable, with greatest frequency around a ZINDO-computed HOMO eigenvalue of -6 eV and LUMO eigenvalue of -1 eV. Notably, this hot spot also appears in the 908 monomer set and the 1759 set. Interestingly, while most top organic photovoltaics are synthesized with a donor-acceptor motif, the presence of one main region of top candidate monomers suggests a donor-donor strategy is also possible.

**Top Monomers from the GA**. While the determination of the most likely monomer candidates for solar applications is the goal of running this genetic algorithm, deciding how to define this group of top monomers was not straight-forward. In our previous work with 129 monomers, we could easily analyze the top 25 monomers. When comparing sets of data with substantially different numbers of monomer, we realized that examining the top X% of the data set was a more meaningful comparison since most of the data comes from a small subset of the monomers as shown in Figure 3 and by the histograms in Figure S44. To determine which percentages to examine, the number of monomers which comprise 25%, 50%, 75%, 80%, 90%, 95%, 98% and 99% were examined and converted to percentages of the total data for comparison between different sized data sets. The results from this are reported in Table 3, and show that up to 90% of the data comes from a small percentage of the total number of monomers. We therefore analyze the top 10% and the top 20% of the monomers as these data sets contain most of the data from a given run of the GA.

**Table 3.** Analysis of the percentage of monomers which comprise the top oligomers reveals that up to 90% of the final candidate oligomers comes from a small fraction of the initial monomers (i.e., ~8-20% of all possible monomers).

| Percentage of Top Candidate Oligomers | Fraction of monomers that comprise top X% of candidate oligomers after 100 generations | | | | | |
|---|---|---|---|---|---|---|
| | 129 | 442 | 611 | 908 | 1235 | 1759 |
| 25% | 2% | 1% | 2% | 1% | 1% | 1% |
| 50% | 5% | 3% | 2% | 4% | 3% | 2% |
| 75% | 8% | 5% | 12% | 6% | 6% | 5% |
| 80% | 10% | 6% | 14% | 7% | 7% | 5% |
| 85% | 12% | 7% | 17% | 9% | 9% | 6% |
| 90% | 18% | 10% | 23% | 11% | 11% | 8% |
| 95% | 34% | 23% | 36% | 27% | 19% | 16% |

Naturally, given the small fraction of monomers comprising top candidate oligomers, a critical question is the structure of these monomers and whether they remain top candidates in larger searches. As a stochastic search, a key question is whether the top results are retained across multiple GA runs. We first compared the list of top 10% and top 20% of monomers after reaching the convergence points in Table 2, and after 100 generations. Both lists were highly correlated, indicating the convergence criteria used are reliable (i.e., the Spearman correlation threshold of 0.5). Next, Table 4 presents a pairwise analysis between top monomers from a smaller search space that are retained in the top monomers from larger data sets. In both the top 10% and 20% of the candidate monomers, an overwhelming number appear in the larger sets of data, even through each run starts with a different random pool of candidates from the given data set. In short, since the main goal of the GA search is to find the top monomer genes, despite the random nature of the algorithm, we find a highly converged set across all search sizes and multiple runs.

**Table 4**. Overlap analysis of the pairwise combination of the data for the top 10% and top 20% of the monomers reveals that most of the candidates that are chosen in the smallest data group (129) are still chosen as important when the search space is increased by more than 10-fold.

|  | Top 10% of Monomers | | | | | | Top 20% of Monomers | | | | | |
|---|---|---|---|---|---|---|---|---|---|---|---|---|
|  | 129 | 442 | 611 | 908 | 1235 | 1759 | 129 | 442 | 611 | 908 | 1235 | 1759 |
| **129** |  | 85% | 64% | 79% | 64% | 71% |  | 59% | 52% | 48% | 52% | 59% |
| **442** |  |  | 70% | 67% | 60% | 60% |  |  | 62% | 56% | 54% | 54% |
| **611** |  |  |  | 47% | 42% | 40% |  |  |  | 36% | 40% | 39% |
| **908** |  |  |  |  | 33% | 48% |  |  |  |  | 31% | 59% |
| **1235** |  |  |  |  |  | 75% |  |  |  |  |  | 65% |

Building on the finding that there is significant overlap in the pairwise analysis of the different sized data sets, we questioned whether there are a group of candidate molecules which appear in the top 10% and the top 20% of *all* data sets examined in this study. Figure 4 illustrates the seven monomers that are present in *all* data sets in both the top 10% and the top 20% of the data. While only three of these species have been synthesized to our knowledge, all seven have previously been cited in papers or patents as potential materials for improved OPV materials. While some may be challenging to synthesize, the retained bi-aromatic ring structure is clearly an important motif, and the presence of hotspots in Figure 3, indicate related analogues with greater synthetic accessibility are likely. We note that none of the top seven monomers are typical electron-poor acceptor motifs, suggesting limits to the conventional donor-acceptor strategy ubiquitous in organic photovoltaic materials.

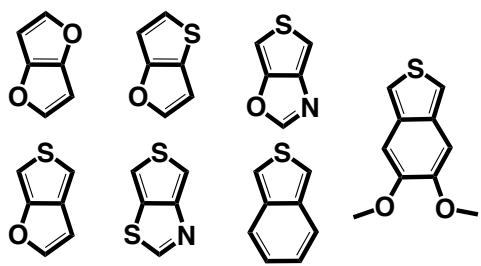

**Figure 4.** The top seven monomers which appear across *all* data sets for the top 10% *and* top 20% of the data.

**Sequence Analysis.** As with the top monomer analysis, we examined the sequences chosen by the GA to determine if particular combinations are more frequent. After examining all the data sets, we determine that each of the sequences is chosen about equally except for AAAA which is surprisingly chosen ~10 times more often compared with the other possible sequences. Calculations of three data sets with hexamers (129, 442 and 908 monomers) indicate that the same bias exists with hexamers towards a homo-oligomer. Interestingly, despite the ubiquitous use of copolymers in experimental investigation of organic electronics, the GA indicates selected homopolymers may still have interesting properties.

**Top Monomer Pairs.**

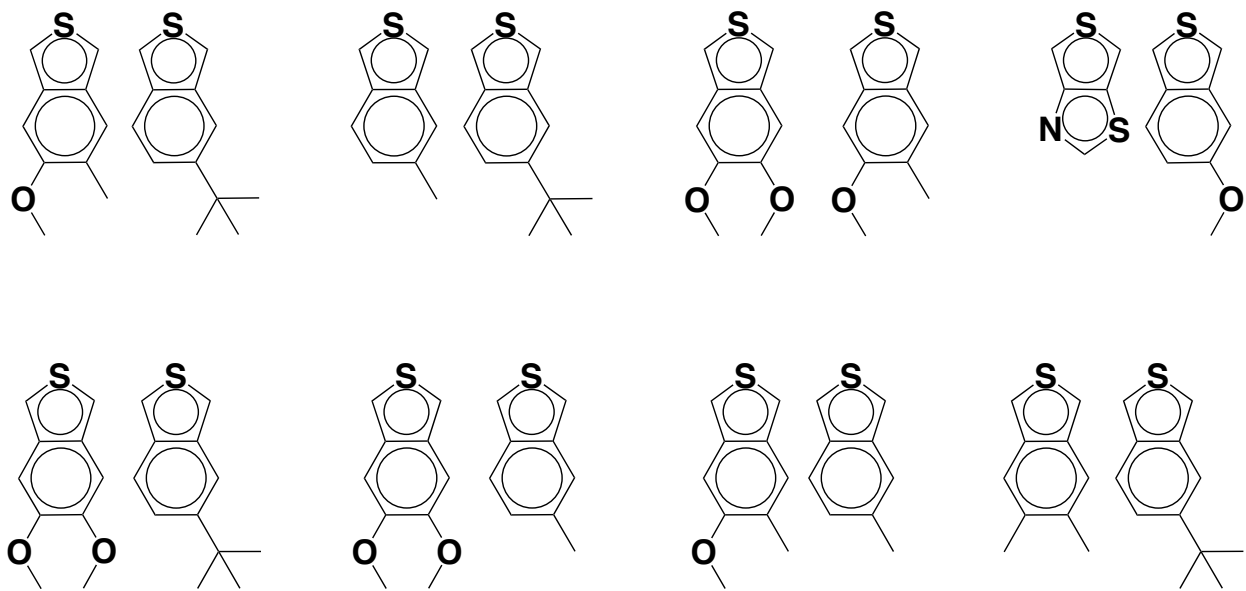

**Figure 5**. Monomer pairs from tetramer GA runs with 442, 611, 908, 1235 and 1759 candidates after 100 generations revealed that the same pairs are chosen most often regardless of size of data set.

In addition to analysis of top monomers chosen, selected monomer pairs were also examined. In each step of the GA, two monomers are chosen to form a co-oligomer and therefore, it is logical to look at the pairs with the top performance and frequency. Data was examined after 100 generations for each of the tetramer runs by selecting each of the two monomers, ensuring a unique combination (e.g., AB = BA) and then counting the frequency of each pair. Figure 5 illustrates the eight monomer pairs that were selected most often across the 442, 611, 908, 1235 and 1759 data sets. The set of data with 129 candidates is excluded from this analysis, since some of the monomers were not present in this smaller set. This data is corroborated with the hexamers from the 442 and 908 runs which show the same set of top monomers. Clearly the isobenzothiophene, a known low-band gap system, is an influential motif, particularly when combined with electron-donating substituents such as alkyl or alkoxy functional groups.

CONCLUSIONS

Genetic algorithm methods are known as efficient techniques for optimizing discrete variables, such as molecular structures. This work shows that, when selecting conjugated oligomers, our GA approach is ~6000-800 times faster than brute force. Since the GA is independent of computational methods used, it can provide rapid filtering of lead compounds using a wide range of electronic structure calculation methods. In short, while we have focused on semiempirical PM6 and ZINDO Hamiltonians in this work, we believe additional performance can be achieved by combining with machine learning methods such as neural networks to predict first principals DFT and other quantum chemical methods. [32-36] Moreover, we find only modest scaling of the number of generations required to converge the set of top candidates up to a search space of ~25 million compounds.

Unfortunately, convergence, as judged by the Spearman rank correlation of the top candidates, is not found for a larger search space of 50 million compounds. Instead, the GA appears stuck in local regions and cannot explore the entire (vast) search space. In future work, we believe a divide-and-conquer approach that partitions large searches into smaller regions, combined with a competition among top candidates will address this problem.

Moreover, we find all searches, regions of "hot spots" (Figure 3) which comprise monomers frequently incorporated in top candidates. This suggests broader searches can perform some level of initial filtering based on these properties. That is, a new monomer found far from a hotspot is unlikely to be among top candidates and can likely be ignored. Predicting the frequency of monomer genes among top candidates will clearly improve the efficiency of the GA search – the presence of such hotspots suggests that for organic electronic materials, statistical and machine learning approaches can rapidly identify interesting new leads.

Molecular space is known to be enormous, but the application of efficient GA search techniques offers great promise for finding optimal and near-optimal targets for a wide range of computationally-driven properties. The techniques outlined here for organic electronic materials can easily be adopted for many other electronic structure properties, from redox potentials and activation energies, to polarity, polarizability and dielectric constants.

**Supporting Information**. Full details of the genetic algorithm implementation, tables and figures of monomer structures, figures of search convergence, and histograms of top monomers.

**Corresponding Author**

* geoffh@pitt.edu

**ACKNOWLEDGMENT** We thank the University of Pittsburgh, including the Center for Energy for support, and NSF (CBET-1404591). GRH thanks Dr. Noel O'Boyle and Prof. Tara Meyer for discussions.